\documentclass[1pt,letterpaper]{article}
\usepackage[margin=1 in]{geometry}
\usepackage[latin1]{inputenc}
\usepackage{amsmath,amsthm}
\usepackage{amsmath,bm}
\usepackage{amsmath,mathtools}
\usepackage{amsfonts}
\usepackage{amssymb}
\usepackage{eurosym}
\usepackage{float}
\usepackage{esint}
\usepackage{xcolor}
\usepackage{titling}
\usepackage{graphicx}
\usepackage{caption}
\usepackage{subfigure}
\usepackage{hhline}

\usepackage[section]{placeins}
\usepackage[pdftex,colorlinks=true,urlcolor=blue,citecolor=brown]{hyperref}

\theoremstyle{remark}

\theoremstyle{definition}

\numberwithin{equation}{section}

\begin{document}

\title{Optical characterization of dyed liquid crystal cells}
\author{ Obeng Appiagyei Addai$^{1}$\thanks{
Email: oaddai@kent.edu, ORCID 0000-0003-0024-7989}, Ruilin Xiao$^{2}$\thanks{
Email: rxiao@kent.edu, ORCID 0000-0001-9001-098X},
Xiaoyu Zheng$^{1}$ \thanks{
Corresponding author, Email: xzheng3@kent.edu, ORCID 0000-0002-3787-7741},
Peter Palffy-Muhoray$^{1,2}$\thanks{
Email: mpalffy@kent.edu, ORCID 0000-0002-9685-5489} \\
\emph{$^1$Department of Mathematical Sciences, Kent State University, OH,
USA }\\
\emph{$^2$Advanced Materials and Liquid Crystal Institute, Kent State
University, OH, USA} }
\maketitle

\begin{abstract}
The guest-host liquid crystal display, first proposed in 1968, relies on
controlling the orientation of dichroic dyes dissolved in a nematic liquid
crystal host. Controlling the orientation of the liquid crystal and of the
dissolved dye with an electric field allows control of the transmittance of
the cell. Knowing the dielectric properties at optical frequencies of the dye and liquid crystal
mixtures is crucial for the optimal design of guest-host liquid crystal
devices. In this work, the dielectric functions of various layers in liquid
crystal cells are described by models obeying the Kramers-Kronig relations:
the Sellmeier equation for transparent layers and causal Gaussian oscillator
model for absorbing layers. We propose a systematic way to accurately model
the dielectric response of each layer by minimizing the sum of squared
differences between the measured transmittance spectrum of a guest-host cell
in the near-UV/vis range and the prediction of the transmittance of the
modeled multilayer structure. By measuring the transmittance for incident
light polarized parallel and perpendicular to the nematic director allows us
to separately characterize the two principal dielectric functions of the
uniaxial sample. Our results show that the causal Gaussian oscillator model
can accurately characterize the dielectric functions of dyes in liquid
crystals.
\end{abstract}

keywords: dielectric functions, dichroic dyes, guest-host liquid crystals, causal Gaussian oscillator model

\section{Introduction}

Dyes are colored since their absorption is wavelength dependent. They have a
wide range of applications, including information displays, such as LCDs.
The first guest-host liquid crystal display was proposed by Heilmeier and
Zanoni in 1968 \cite{Heilmeier1968}, where dichroic dyes are dissolved in
nematic liquid crystals (NLCs). The transmittance of the NLC cell is
controlled by an electric field which can orient the LC molecules, which in
turn orient the dichroic dye molecules. The optical response depends on polarizabilities of the constituents, hence their dielectric tensors at optical frequencies. Many guest-host liquid crystal
systems have been subsequently proposed and realized; various aspects of
these systems have been studied and reported, as in early review papers by
Cox in 1979 \cite{Cox1979} and by Scheffer in 1983 \cite{Scheffer1983}, and
more recently by Sims in 2016 \cite{Sims2016}.

Modeling the dielectric function of dyes in LCs at optical frequencies can be a powerful tool for
the interpretation of measured transmittance spectra and for predicting the
transmittance and reflectance as function of wavelength. Characterization of
the dielectric function of dyes in solutions is a challenging task in
general, as dyes molecules often show the tendency to associate causing the
transmittance to depend nonlinearly on concentration and absorption peaks to
shift when mixing two dyes \cite{NIST1958}. However, at low
concentrations, one can assume the linear response.

An overview of most popular models for dielectric functions and their
applicabilities can be found in \cite{Hilfiker2018}. Among those models, the
classical Lorentz model assumes that the motion of the electrons, bound to
atomic nuclei, can be modeled as a simple damped harmonic oscillator,
subject to the external electromagnetic fields. The dielectric function $%
\varepsilon $ can then be expressed as (e.g. \cite{Born33}) 
\begin{equation}
\varepsilon ^{LO}(\omega )=\varepsilon _{\infty }+\sum_{k=1}^{N}\frac{%
A_{k}\omega _{p}^{2}}{\omega _{0k}^{2}-\omega ^{2}-i\gamma _{k}\omega },
\label{eq_LO}
\end{equation}%
where $\varepsilon _{\infty }$ is the high frequency dielectric constant, $%
\omega _{p}$ is the plasma frequency, $A_{k}$ is the strength of the $k$th
oscillator with resonant frequency $\omega _{0k}$ and the damping
coefficient $\gamma _{k}$. The damping coefficient $\gamma _{k}$ is
proportional to the width of the absorption peak. Taking the limit of of $%
\gamma _{k}\rightarrow 0$ in Eq.~(\ref{eq_LO}) gives the Sellmeier equation, 
\begin{equation}
\varepsilon ^{S}(\omega )=\varepsilon _{\infty }+\sum_{k=1}^{N}\frac{%
A_{k}\omega _{p}^{2}}{\omega _{0k}^{2}-\omega ^{2}},  \label{eq_S}
\end{equation}%
which is often used to model the dispersion relations for transparent
materials away from the absorption peaks. The Lorentz oscillator model
satisfies the Kramers-Kronig (KK) causality relations. Some authors \cite%
{Taqatqa2007, Andam2021} have used Lorentz oscillator models for dispersion
relations for azo-dyes. The Lorentz oscillator model can produce realistic
absorption frequencies and amplitudes. However, as the Lorentz oscillator
model tends to predict slower decays, the fitting to the
absorption data is poor away from the absorbance peaks \cite{Andam2021,
Duorovic2017}.

In 1992, Brendel and Bormann (BB) proposed a dielectric function model for
amorphous solids in the infrared (IR) range \cite{Brendel&B92}. It is
written as a convolution of a Gaussian function with the Lorentz oscillator 
\begin{equation}
\varepsilon ^{BB}(\omega )=\varepsilon _{\infty }+\sum_{k=1}^{N}\chi
_{k}^{BB}(\omega ),
\end{equation}%
where%
\begin{equation}
\chi _{k}^{BB}(\omega )=\frac{1}{\sqrt{2\pi }\sigma _{k}}\int_{-\infty
}^{\infty }\exp \left( -\frac{(x-\omega _{0k})^{2}}{2\sigma _{k}^{2}}\right) 
\frac{A_{k}\omega _{p}^{2}}{(x^{2}-\omega ^{2})-i\omega \gamma _{k}}dx.
\end{equation}%
It was interpreted as the superposition of damped harmonic oscillators whose
resonance frequencies follow a Gaussian distribution centered at $\omega
_{0k}$ with standard deviation $\sigma _{k}$. Some authors \cite%
{Duorovic2017, Ni2010, Darby2016} used the Brendel and Bormann model,
sometimes referred to as Voigt model, for modeling the dielectric functions
of dyes and showed improved accuracy over the Lorentz oscillators.
Subsequently it was shown that the BB model is not causal as it fails to
satisfy $\chi _{k}^{BB}(\omega )\neq \overline{\chi _{k}^{BB}(-\omega )}$ in
addition to a divergent behavior near $\omega =0$ \cite{Coimbra2018}.

Meneses \textit{et. al.} \cite{Echegut2006} in 2006 proposed the causal
Gaussian oscillator model, to model the dielectric function of binary lead
silicate glasses, 
\begin{equation}
\varepsilon ^{CG}(\omega )=\varepsilon _{\infty }+\sum_{k=1}^{N}\chi
_{k}^{CG}(\omega ),  \label{eq_CG_1}
\end{equation}%
where the imaginary part of $\chi _{k}^{CG}$ is given by%
\begin{equation}
\chi _{k}^{CG\prime \prime }(\omega )=A_{k}\left[ \exp \left( -\frac{(\omega
-\omega _{0k})^{2}}{2\sigma _{k}^{2}}\right) -\exp \left( -\frac{(\omega
+\omega _{0k})^{2}}{2\sigma _{k}^{2}}\right) \right] ,  \label{eq_CG_2}
\end{equation}%
and the real part, which guarantees satisfying the KK relations, is given in
terms of the Dawson function $D(x)$ as 
\begin{equation}
\chi _{k}^{CG\prime }(\omega )=\frac{2A_{k}}{\sqrt{\pi }}\left[ D\left( 
\frac{\omega +\omega _{0k}}{\sqrt{2}\sigma _{k}}\right) -D\left( \frac{%
\omega -\omega _{0k}}{\sqrt{2}\sigma _{k}}\right) \right] ,  \label{eq_CG_3}
\end{equation}%
where $D(x)=e^{-x^{2}}\int_{0}^{x}e^{t^{2}}dt$. The causal Gaussian
oscillator model can be obtained by removing the divergent factor from the
BB oscillator model and setting $\gamma _{k}=0$. It has been used to characterize silica glass in the far IR region \cite%
{Kitamura2007}, and Titanium dioxide from near-UV to near-IR \cite{Huang2013}.   

\textcolor{black}{We point out that in Ref.~\cite{Goda2013}, the authors used the renormalized ellipsometry to determine the device parameters and tilt angles of a guest-host LC system. In particular, they have used a combined Cauchy equation and Gaussian model for the complex refractive indices of a guest-host LC system. Their assumed dielectric function does not obey the KK relations.  Nonetheless, they showed that the extinction coefficient of the guest-host LC system can be characterized by a Gaussian function. }

In another vein, calculations based on density functional theory or
molecular dynamics were carried out to predict the polarizabilities of dyes 
\cite{Duorovic2017} and the alignment of dyes in LCs \cite{Sims2015,
Fegan2018}. These calculations require knowledge of the specific molecular
structures to find the minimum potentials and their frequency dependence.

In this work, the characterization of the dielectric functions of our
samples is solely based on the transmittance spectra, without the knowledge
of their molecular structures. Specifically, we model the dielectric
function of each layer in our multilayer LC samples in the near-UV/vis range
by either the Sellmeier equation, if they appear transparent or by the
causal Gaussian oscillator model, if they have absorption. We show that the
dielectric function of the dye and liquid crystal mixture can be accurately
described by the causal Gaussian oscillator model.

The rest of the paper is organized as follows. In section \ref{Sec_MathModel}%
, we present the mathematical models and methods which lead to the
determination of the parameters in the dielectric function models. Section %
\ref{Sec_Expt} describes our experimental setup for obtaining the
transmittance spectra of our samples. In section \ref{Sec_Res}, we show the
characterization results of dielectric functions of all components in our
guest-host LC sample. We give our conclusions in section \ref{Sec_Con}.

\section{Mathematical Models and Methods}

\label{Sec_MathModel}

In this section, we first briefly discuss the eigenvalue problem associated
with light propagation in a lossy uniaxial material, explain the setup of
our experiments and discuss the validity of applying the averaged
transmittance model of a multilayer structure to our problem. Then we
present the least squares approach used to determine the parameters in the
dielectric function models where we minimize the sum of the squared
differences between the experimental measurements and the results of
theoretical models.

A uniform uniaxial material is birefringent, and the polarization of plane
polarized light will, in general, change as it propagates through the
material unless the polarization is along one of two eigen directions
associated with the wave propagation direction. The eigenvalue problem in
term of the displacement field $\mathbf{D}$, derived from Maxwell equation,
is given by (cf. \cite{Born33})%
\begin{equation}
(k^{2}\mathbf{I}-\mathbf{kk})\varepsilon _{r}^{-1}\mathbf{D}=k_{0}^{2}%
\mathbf{D},
\end{equation}%
where $\mathbf{I}$ is the identity tensor, $\mathbf{k}=k^{\prime }\mathbf{%
\hat{k}}^{\prime }+ik^{\prime \prime }\mathbf{\hat{k}}^{\prime \prime }$ is
the complex wave vector, $k^{2}=\mathbf{k}\cdot \mathbf{k}$ $=k^{\prime
2}-k^{\prime \prime 2}+2ik^{\prime }k^{\prime \prime }(\mathbf{\hat{k}}%
^{\prime }\cdot \mathbf{\hat{k}}^{\prime \prime })$ is the dot product, $%
\mathbf{kk}$ is the tensor product, and $k_{0}^{2}=\omega ^{2}\mu_0
\varepsilon _{0}$, where $\mu_0$ is the permeability, $\varepsilon _{0}$ is
permittivity of free space, and $\omega $ is the angular frequency. For
uniaxial absorbing materials, the complex relative dielectric tensor can be
written as 
\begin{equation}
\varepsilon _{r}=\varepsilon _{\perp }\mathbf{I}+(\varepsilon _{\parallel
}-\varepsilon _{\perp })\mathbf{\hat{n}\hat{n},}
\end{equation}%
where $\varepsilon _{\parallel }$ is the dielectric constant for fields
parallel to the symmetry axis $\mathbf{\hat{n}}$, and $%
\varepsilon _{\perp }$ is the dielectric constant for fields in the plane
perpendicular to $\mathbf{\hat{n}}$. In our LC samples, the symmetry axis $%
\mathbf{\hat{n}}$ coincides with the director of the nematic liquid crystal.

Given the directions of $\mathbf{\hat{k}}^{\prime }$ and $\mathbf{\hat{k}%
^{\prime \prime }}$ and $\varepsilon _{r}$, one can solve for the
eigenvalues $k^{\prime }$, $k^{\prime \prime }$, and for the corresponding
eigenvector $\mathbf{D}$. One eigen-pair is given by 
\begin{equation}
k_{1}^{2}=k_{0}^{2}\varepsilon _{\perp },\mathbf{D}_{1}\parallel \mathbf{%
\hat{n}}\times \mathbf{k}_{1},  \label{eqn_firsteig}
\end{equation}%
while the other is given, implicitly, by 
\begin{equation}
k_{2}^{2}\varepsilon _{\perp }+\Delta \varepsilon (\mathbf{\hat{n}\cdot k}%
_{2})^{2}=k_{0}^{2}\varepsilon _{\parallel }\varepsilon _{\perp },
\label{eqn_secondeig_1}
\end{equation}%
and 
\begin{equation}
\mathbf{D}_{2}\parallel \mathbf{k}_{2}\times (\mathbf{\hat{n}}\times \mathbf{%
k}_{2}).  \label{eqn_secondeig_2}
\end{equation}%
Similar results can be found in \cite{Dineiro2007}.

Suppose light propagating in an isotropic lossless medium is normally
incident on a nematic LC cell where the director $\mathbf{\hat{n}}$ is along
the $\mathbf{\hat{x}}$ direction. Boundary conditions require that, in the
cell, $\mathbf{\hat{k}}=\mathbf{\hat{k}}^{\prime }=\mathbf{\hat{k}}^{\prime
\prime }=\mathbf{\hat{z}}$. If the polarization of plane polarized
incident light, and hence the direction $\mathbf{\hat{E}}$ of the electric
field of the incident light is perpendicular to $\mathbf{\hat{n}}$, we have,
by Eq.~\eqref{eqn_firsteig}, 
\begin{equation}
k_{1}^{2}=k_{0}^{2}\varepsilon _{\perp },\mathbf{D}_{1}\parallel \mathbf{E}%
_{1}\parallel \mathbf{\ \hat{y}.}
\end{equation}%
That is, the polarization of the light will remain in the $\mathbf{\hat{y}}$%
-direction, and the wave number in the sample is only dependent on $%
\varepsilon _{\perp }$. If the linear polarization is such that $\mathbf{%
\hat{E}}$ is parallel to $\mathbf{\hat{n}}$, we obtain from Eqs.~%
\eqref{eqn_secondeig_1}-\eqref{eqn_secondeig_2}, 
\begin{equation}
k_{2}^{2}=k_{0}^{2}\varepsilon _{\parallel },\mathbf{D}_{2}\parallel \mathbf{%
E}_{2}\parallel \mathbf{\hat{x}.}
\end{equation}%
That is, the polarization of the light will remain in the $\mathbf{\hat{x}}$%
-direction, and the wave number in the sample is only dependent on $%
\varepsilon _{\parallel }$. Those correspond to the two cases depicted in
Fig.~\ref{fig:exp_setup}, and indicate the geometries considered in this
work.

In these two scenarios, we can then use the following recursive relations to
calculate the transmission and reflection coefficients for light propagation
through an isotropic multilayer structure at normal incidence, with light
incident from medium 1 and exiting to medium $m$ (see, e.g. \cite%
{Orfanidis2002}), 
\begin{align}
t_{1:m}& =\frac{t_{12}t_{2:m}e^{ik_{0}n_{2}d_{2}}}{%
1+r_{12}r_{2:m}e^{2ik_{0}n_{2}d_{2}}},  \label{eqn_t_1} \\
r_{1:m}& =\frac{r_{12}+r_{2:m}e^{2ik_{0}n_{2}d_{2}}}{%
1+r_{12}r_{2:m}e^{2ik_{0}n_{2}d_{2}}},  \label{eqn_t_2}
\end{align}%
where 
\begin{align}
r_{ij}& =\frac{n_{i}-n_{j}}{n_{i}+n_{j}},t_{ij}=\frac{2n_{i}}{n_{i}+n_{j}},
\\
n_{i}& =\sqrt{\varepsilon _{i}},{i=1,\cdots ,m}.
\end{align}%
Here $n_{i}$ is the complex refractive index whose real and imaginary parts
are referred to as the index of refraction and the extinction coefficient,
respectively. The transmittance and reflectance of the multilayer structure,
with $n_{1}=n_{m}$, can be readily obtained from
\begin{equation}
T_{1:m}=|t_{1:m}|^{2},R_{1:m}=|r_{1:m}|^{2}.  \label{eqn_TR}
\end{equation}%
A general result for transmittance and reflectance of anisotropic absorbing
multilayer structure can be found in \cite{Papousek2001}. We note that a
decrease in the transmittance of a multilayer structure can be due to
absorption in the layers and/or to reflections at the interfaces. In the
case of absorption, the refractive index has a nonzero imaginary part,
originating in the imaginary part of $\varepsilon $, and in the case of
reflections, there is a mismatch of the refractive indices of adjacent
layers.

The transmittance and reflectance from Eq.~\eqref{eqn_TR} of a thick ($d\gg
\lambda )$ single- or multilayer structure can show large and rapid
oscillations as function wavelength due to interference and multiple
reflections. These become increasingly difficult to observe experimentally
as the sample thickness increases, due to the finite resolution of spectrometers
and sample inhomogeneities. In this limit, ignoring multireflection
interference, or, equivalently, by assuming random phase, the average
transmittance $\mathcal{T}_{1:m}$ and reflectance $\mathcal{R}_{1:m}$ for a
multilayer structure at normal incidence are given by%
\begin{eqnarray}
\mathcal{T}_{1:m} &=&\frac{\mathcal{T}_{12}\mathcal{T}_{2:m}e^{-2k_{0}\text{%
Im}(n_{2})d_{2}}}{1-\mathcal{R}_{21}\mathcal{R}_{2:m}e^{-4k_{0}\text{Im}%
(n_{2})d_{2}}},  \label{AvgTrans} \\
\mathcal{R}_{1:m} &=&\mathcal{R}_{12}+\frac{\mathcal{T}_{12}\mathcal{R}_{2:m}%
\mathcal{T}_{21}e^{-4k_{0}\text{Im}(n_{2})d_{2}}}{1-\mathcal{R}_{21}\mathcal{%
R}_{2:m}e^{-4k_{0}\text{Im}(n_{2})d_{2}}},  \label{AvgRef}
\end{eqnarray}%
where 
\begin{equation}
\mathcal{R}_{ij}=\left\vert \frac{n_{i}-n_{j}}{n_{i}+n_{j}}\right\vert ^{2},%
\mathcal{T}_{ij}=\left\vert \frac{2n_{i}}{n_{i}+n_{j}}\right\vert ^{2}.
\label{eq_Ave4}
\end{equation}%
A similar expression for a single etalon with $m=3$ can be found in \cite%
{Olaf2015}. In our study, we have set $n_{1}=n_{m}=n_{air}=1$.

By assuming a specific form for the dielectric function of each layer, the
parameters in the dielectric functions can be found by minimizing the sum of
the squares of the errors between the theoretical transmittance of a
multilayer structure and the experimental data. Specifically, the
minimization problem can be written as 
\begin{equation}
\min \sum_{i=1}^{M}[\left( y_{\parallel }(\omega _{i})-\mathcal{T}%
_{1:m}(\omega _{i};n_{1\parallel },\cdots ,n_{m\parallel })\right)
^{2}+\left( y_{\perp }(\omega _{i})-\mathcal{T}_{1:m}(\omega _{i};n_{1\perp
},\cdots ,n_{m\perp })\right) ^{2}],  \label{eq_min}
\end{equation}%
where $M$ is the total number of data points, $y_{\parallel }(\omega _{i})$
and $y_{\perp }(\omega _{i})$ are the locally averaged transmittance data at 
$\omega _{i}$ when the polarization of the incident light is parallel and
perpendicular to the LC director, respectively. The two principal refractive
indices of each layer are given by 
\begin{equation}
n_{j\parallel }=\sqrt{\varepsilon _{j\parallel }(\varepsilon _{\infty
\parallel }^{j},{\bar{A}_{\parallel }^{j},\bar{\omega}_{0}^{j},\bar{\sigma}}%
^{j})},n_{j\perp }=\sqrt{\varepsilon _{j\perp }({\varepsilon _{\infty \perp
}^{j},\bar{A}_{\perp }^{j},\bar{\omega}_{0}^{j},\bar{\sigma}}^{j})}%
,j=1,\cdots ,m,
\end{equation}%
and the dielectric functions $\varepsilon _{j\parallel }$ and $\varepsilon
_{j\perp }$ are modeled either as a Sellmeier equation in Eq.~\eqref{eq_S},
or as a causal Gaussian oscillator model in Eqs.~\eqref{eq_CG_1}-%
\eqref{eq_CG_3}, where $\varepsilon _{\infty \parallel }^{j},\varepsilon
_{\infty \perp }^{j},{\bar{A}_{\parallel }^{j},\bar{A}_{\perp }^{j},\bar{%
\omega}_{0}^{j},\bar{\sigma}}^{j}$ are the fitting parameters. The lengths
of the vector parameters ${\bar{A}_{\parallel }^{j},\bar{A}_{\perp }^{j},%
\bar{\omega}_{0}^{j},\bar{\sigma}}^{j}$are equal to the number of
oscillators. We have assumed that the peak locations $\bar{\omega}_{0}^{j}$
and their widths $\bar{\sigma}^{j}$ along the two principal directions are
the same, but the amplitudes $\bar{A}_{\parallel }^{j},\bar{A}_{\perp }^{j}$
are different, as suggested by the experimental data.

We note that one can't model all the layers collectively as an effective
single layer, which disregards the arrangement of the layers, since, in
general, the transmittance of the multilayer structure may sensitively depends
on the order of the layers. It is also impossible to determine the
refractive indices of all layers simultaneously from a single transmittance
measurement of all layers, as the solutions won't be unique. Therefore,
ideally, we should determine the dielectric response of each layer
separately, but this is often not practically viable. However, we can
measure, for example, the transmittance of a single glass layer first, and
then that of a single glass layer coated with indium tin oxide (ITO) layer,
and next that of a polyimide (PI) alignment layer on top of the glass and
ITO layers. These three measurements enable us to determine the dielectric
properties of glass, ITO, and PI iteratively, one at a time. To determine
the dielectric properties of dye-LC mixtures, we assume the effect of a dye
guest on the dielectric response of the mixture is an additive contribution
to that of the pure LC host. With this in mind, we first use the
transmittance spectrum of pure LC in the cell to determine the dielectric
function of LC host, and then use the transmittance spectrum of dye-LC
mixture in the cell to determine the contribution of dye to the dielectric
function of dye-LC mixture.

We have implemented the Levenberg$-$Marquardt algorithm to solve the above
minimization problems. \textcolor{black}{It is important to note that the function to be minimized, Eq.~\eqref{eq_min}, has multiple local minima. To obtain the most faithful fitting results, we have carefully selected the initial parameters which give a crude agreement of the experimental and model transmittances. Specifically, the absorption peak positions $\omega_{0k}$ were chosen to be in proximity of the dips of the transmittance spectra.} The number of oscillators used in modeling each layer
is determined by whether or not the addition of one more oscillator improves
the fitting significantly. The goodness of fit is assessed from maximum pointwise
errors, rather than an averaged error. 

\section{Experiment}

\label{Sec_Expt}

Our liquid crystal cells typically use two glass substrates with each
substrate coated with a conductive ITO layer enabling the application of an
electric field across the cell. The PI alignment layer, next to the ITO
layer, is buffed antiparallel on the two substrates
to impose the homogeneous boundary conditions on the nematic directors. The
thicknesses of the glass, ITO and PI layers are approximately $1$mm, $100$
nm and $100$ nm respectively. The thickness of the cell gap is approximately 
$20\mu m$. The empty cells are filled with pure LCs or with dye-LC mixtures,
with $1\%$ or $0.5\%$ dye concentrations.

\begin{figure}[h]
\centering
(a)\includegraphics[scale=0.5]{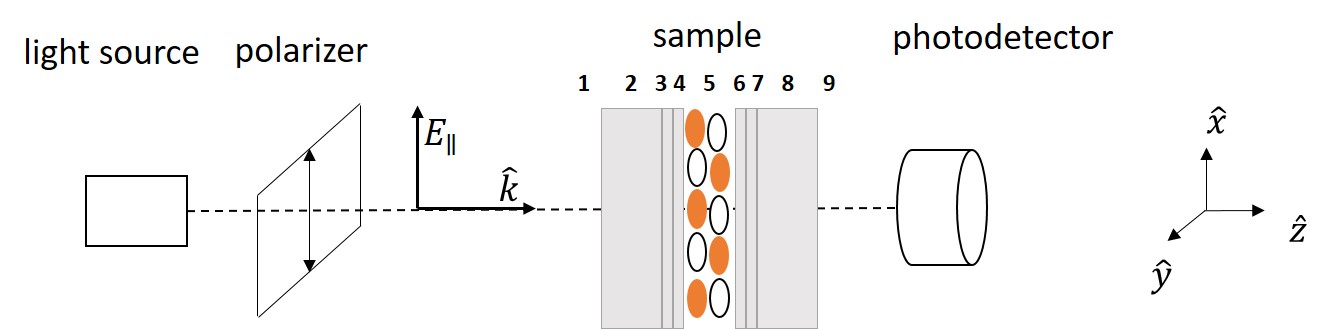} (b)%
\includegraphics[scale=0.5]{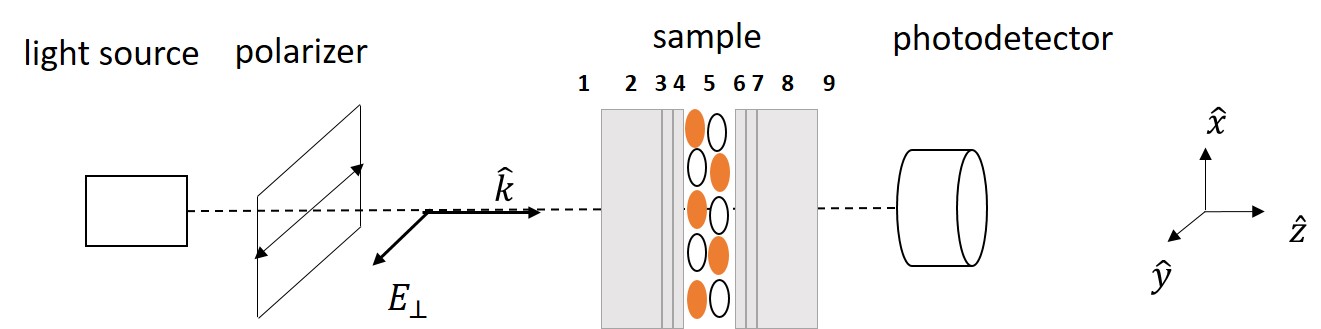}
\caption{Schematic of the experimental setup. Light propagating in the $\mathbf{\hat{z}}-$ direction is normally incident on
the cell. The LC director $\mathbf{\hat{n}}$
is in the $\mathbf{\hat{x}}-$ direction. The transmittance is measured
when the polarization of the light is (a) parallel to the $\hat{\mathbf{n}}$
, and (b) perpendicular to $\hat{\mathbf{n}}$. The sample between the
polarizer and the photodetector represents the LC cell where a dye-LC mixture is contained in two glass(2,8) substrates coated with
ITO (3,7) and PI (4,6) layers. Layers labeled 1 and 9 are air. The sample may also be a single glass, glass coated with ITO, glass coated with ITO
and PI for separate measurements.}
\label{fig:exp_setup}
\end{figure}

As discussed in Section \ref{Sec_MathModel}, we separately measured the
transmittance spectra of the following samples: a single glass, glass with
ITO layer, glass with ITO and PI layers, empty cell, pure LC in the cell,
and dye-LC mixtures in the cell. To measure the transmittance spectra of
each sample, we use a spectrometer with a polarizer. The
sample is aligned so that light is normal incident on the cell. The
polarizer is oriented so that light is polarized parallel, as shown in Fig.~%
\ref{fig:exp_setup}(a), or perpendicular, in Fig. \ref{fig:exp_setup}(b), to
the nematic director.

The transmission spectra are measured using an Ocean Optics HR400CG-UV-NIR
High-Resolution Spectrometer with resolution $\sim $ $0.2-0.3$ nm with an
Ocean Insight DH-2000-BAL light source covering the range of $300$ nm to $%
900 $ nm.

Since the resolution of the spectrometer is high and sample is thin, the
transmittance shows some oscillations due to multiple reflections and
interference. As it is preferable to work with a smooth spectra for fitting
purposes, local averaging of the transmittance was carried out before the
fitting procedure. Measurement errors were estimated by comparing repeated
spectral measurements on single cells. Variations due to differences in cell
properties were estimated by comparing the spectra of six empty cells. The
combined variations gave an estimated uncertainty of approximately $5\%$ in
the wavelength range $300$ to $900$ nm.

\section{Results}

\label{Sec_Res}In this section, we first characterize the dielectric
function of each substrate layer using the causal Gaussian oscillator model
for glass and the Sellmeier equation for ITO and PI in section \ref%
{sec_substrate}. Knowing the refractive indices of substrate layers, we then
proceed to characterize the dielectric function of the pure LC and two
dye-doped liquid crystal mixtures in section \ref{sec_LC_dye}.

\subsection{Dielectric functions of the substrate layers}

\label{sec_substrate}

The transmittance spectra of the substrate layers: glass only, glass with
ITO, and glass with ITO and PI, are shown in Fig.~\ref{fig:glassITOPI}(a).
The transmittance spectra of the three layers do not show significant
polarization dependence, thus we regard all substrate layers as optically
isotropic.

Our strategy is as follows: we start with the characterization of the
dielectric function of glass with the transmittance spectrum of a single
slab of glass. Next, we obtain the dielectric function of ITO by fitting the
transmittance spectrum of glass with ITO, given we have already known the
refractive index of glass. Lastly, the dielectric function of the PI is
found by fitting the transmittance spectrum of glass with ITO and PI layers,
given the dielectric functions of glass and ITO. The results are shown in
detail below.

\subsubsection{glass}

\label{sec:glass}

Glass is generally taken to be transparent in the visible range. The
dispersion relation was generally modeled with Sellmeier equation in \cite%
{Tatian1984,Ghosh1997}.

In this work, the dielectric function of glass is characterized solely based
on the transmittance spectrum of a single glass slab in the range of $300-900
$ nm. In Eqs.~\eqref{AvgTrans}-\eqref{eq_Ave4}, we set $m=3,n_{1}=n_{3}=1$,
and solve the minimization problem Eq.~\eqref{eq_min} to find $%
n_{2}=n_{glass}$. We first attempted to use the Sellmeier equation for $%
\varepsilon _{glass}$. Here, the one-term Sellmeier equation was fitted with
parameters $\varepsilon _{\infty }=1.5302,A=0.6001,\omega _{0}=2\pi
c/309.19nm$, and the predicted transmittance of the 3-layer structure is
shown as the green dot-dashed curve in Fig.~\ref{fig:glassITOPI}(b). We
observed that the Sellmeier equation cannot accurately capture the sharp
corner in the transmittance spectrum near $300$ nm due to the nature of
polynomial decay away from from the absorption peak; it also fails to model
the observed slow decrease of transmittance between about $600$ and $900$
nm. In addition, since the peak of the oscillator is located at $309$ nm,
the dielectric constant diverges near the peak, which is nonphysical.

We therefore used the causal Gaussian oscillator model for $\varepsilon
_{glass}$. We have used two such oscillators, with one oscillator peak at $%
195$ nm for the strong absorption in the near-UV and one at $987.31$ nm for the slow
decrease in the transmittance over visible to near-IR. The causal Gaussian
oscillator model produces excellent agreement between the theoretical
transmittance of the 3-layer structure and the experimental data as shown in
Fig.~\ref{fig:glassITOPI}(b). \textcolor{black}{The average relative error in the range of $400-900$ nm is $0.8\%.$} The real and imaginary parts of the refractive
indices of glass, calculated via the causal Gaussian oscillator model with
parameters given in Table \ref{tab:glassITOPI}, are plotted in Fig.~\ref%
{fig:RIglassITOPI}. The real part of the refractive index of the glass is
about $1.53-1.54$. \textcolor{black}{A more sophisticated model combining Sellmeier and causal Gaussian can be used \cite{Cushman2016} to produce larger variation in the refractive indices with respect to wavelength. However additional information would be needed to decide which model is preferable, hence we do not pursue this topic in this work.} The refractive index of the glass given by the
causal Gaussian oscillator model is used for the subsequent
characterizations of dielectric functions of the remaining layers of the LC
cell.

\subsubsection{ITO layer}

\label{sec:ITO}

For modeling of the dielectric function of the ITO, in the literature,
either a combination of a single Drude oscillator for free carriers and
double Lorentz oscillators for strong absorption in the near-UV \cite{Losurdo2002},
or a combined Drude, causal Gaussian oscillators and a Cauchy term \cite%
{Cleary2018} were used.

\begin{figure}[thb]
\centering
(a)\includegraphics[width=.4\linewidth]{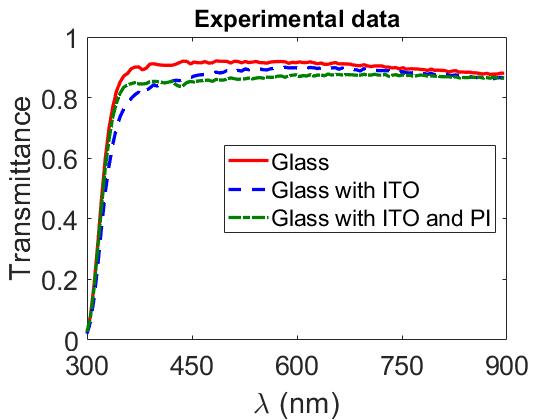} (b)%
\includegraphics[width=.4\linewidth]{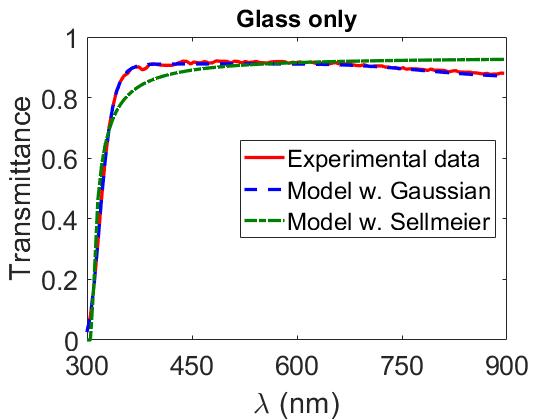} \newline
\newline
(c)\includegraphics[width=.4\linewidth]{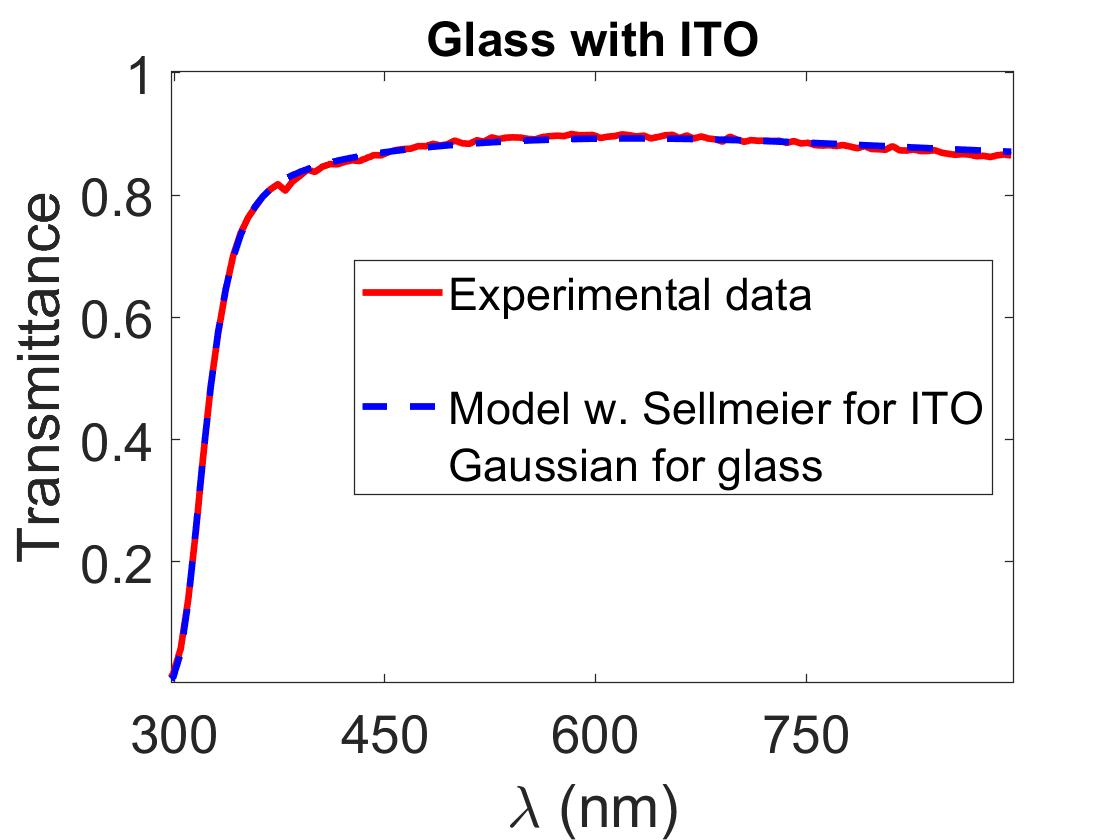} (d)%
\includegraphics[width=.4\linewidth]{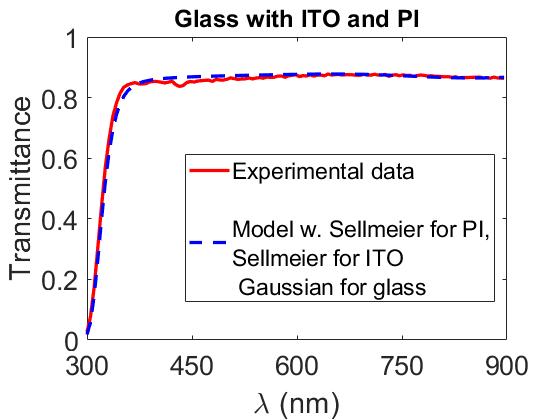} 
\caption{ Transmittance of substrate layers. (a) shows the 
averaged experimental transmittance spectra of glass only, glass with ITO,
and glass with ITO and PI. (b, c, d) show the averaged
experimental data and theoretical prediction of multilayer structures using
dielectric functions of substrate layers with fitting parameters in Table 
\protect\ref{tab:glassITOPI}, for glass only, glass coated with ITO, and
glass coated with ITO and PI, respectively. 
\label{fig:glassITOPI}}
\end{figure}

Here, the dielectric function of ITO is characterized based on the
transmittance spectrum of glass coated with an ITO layer in the range of $%
300-900$ nm. In Eqs.~\eqref{AvgTrans}$-$\eqref{eq_Ave4}, we set $%
m=4,n_{1}=n_{4}=1$, and $n_{2}=n_{glass}^{cG}$, which is obtained from the
causal Gaussian oscillator model in Sec.~\ref{sec:glass}. We then solve the
minimization problem Eq.~\eqref{eq_min} to find $n_{3}=n_{ITO}$. By
comparing the transmittance of glass with ITO with that of glass with ITO
and PI in Fig.~\ref{fig:glassITOPI}(a), we observe that the transmittance of
the 5- layer structure in the near-UV is higher than that of the 4-layer structure.
Absorption in the near-UV from the ITO or from the PI won't be able to produce such
a spectrum. This increase in the transmittance with more layers can only be
explained by index matching. Therefore, we model the dielectric functions of
both the ITO layer and PI layer with the Sellmeier equation. For the ITO,
the Sellmeier equation with two peaks, one at $275.01$ nm and the other at $1000.01$ nm, produces excellent
agreement between the theoretical prediction of transmittance of the 4-layer
structure and the data, as shown in Fig. \ref{fig:glassITOPI}(c), \textcolor{black}{with the average relative error $1\%$ in the range of $400-900$nm}. Fig.~\ref%
{fig:RIglassITOPI}(b) shows the refractive index of ITO calculated from the
Sellmeier equation with parameters given in Table~\ref{tab:glassITOPI}. In
the range of $350-900$ nm, the refractive index for ITO decreases from $2.2$
to $1.6$, which is consistent with that in the literature \cite{Konig2014,
Pan2020}.

\subsubsection{PI layer}

\label{sec:PI}

Buffed polyimide polymer layers can be optically anisotropic, with an
average refractive index in the visible of about $1.5-1.8$ depending on
chemical structure and morphology \cite{yang2007, Cole2013}. Our
cells do not show appreciable birefringence, so we model the PI layer with a
single principal dielectric function.

\begin{figure}[htb]
\centering
(a)\includegraphics[width=.4\linewidth]{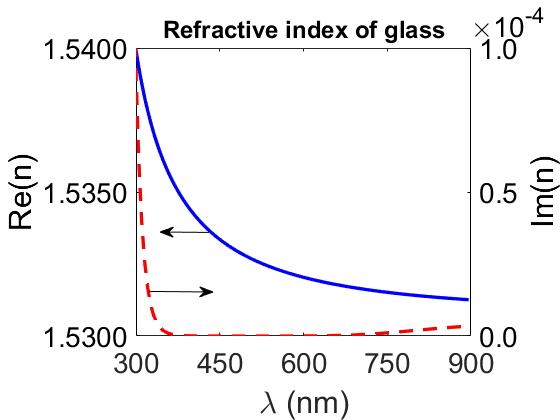}\quad (b)%
\includegraphics[width=.4\linewidth]{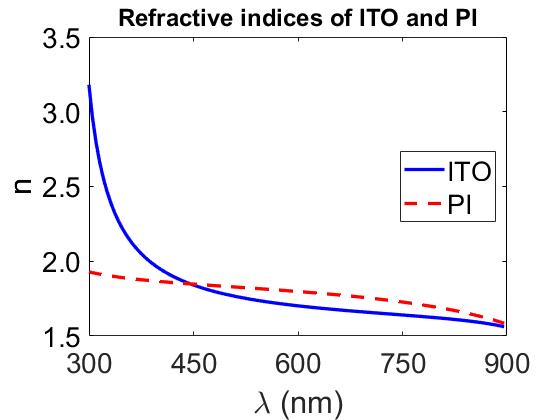}
\caption{(a) Refractive index of the glass substrate. (b) Refractive indices
of the ITO and PI layers. }
\label{fig:RIglassITOPI}
\end{figure}

The dielectric function of PI is determined using the transmittance spectrum
of glass with ITO and PI layers from $300-900$ nm. In Eqs.~\eqref{AvgTrans}$-$%
\eqref{eq_Ave4}, we set $m=5,n_{1}=n_{5}=1$, $n_{2}=n_{glass}^{cG}$ obtained
from Sec.~\ref{sec:glass} and $n_{3}=n_{ITO}^{S}$ obtained from Sec.~\ref%
{sec:ITO}. We then solve the minimization problem Eq.~\eqref{eq_min} to find 
$n_{4}=n_{PI}$. As discussed in \ref{sec:PI}, we model the dielectric
function of PI with the Sellmeier equation. A two-term Sellmeier equation is
fitted with one peak located at $199.32$ nm, and the other in the IR at $1135.22$
nm. \textcolor{black}{The corresponding theoretical transmittance of the 5-layer structure
shows reasonable overall agreement with experimental data, with slightly
larger discrepancy in the $300-500$ nm region, as shown in Fig.~\ref%
{fig:glassITOPI}(d). The average relative error in the range of $400-900$nm is $1\%$.  Surprisingly, the contribution of the PI layer to the transmittance varies dramatically from one piece of substrate to another.  Therefore, we only aim to get a reasonable fitting here. } Fig.~\ref{fig:RIglassITOPI}(b) shows the
refractive index of PI calculated from the Sellmeier equation with
parameters given in Table~\ref{tab:glassITOPI}. The refractive index of PI
decreases from $1.95$ to $1.54$ in the $350$ to $900$ nm range, in agreement
with reported values in \cite{French2009, Kumar2011, Huang2016}.

We note that the fitting parameters of the Sellmeier equations for the ITO
and PI layers are independent of the thicknesses of the ITO and PI layers,
since we have considered the averaged transmittance which depends on the
values of the (real) refractive indices but not on the cell thickness.

\begin{figure}[htb]
\centering
\includegraphics[width=.4\linewidth]{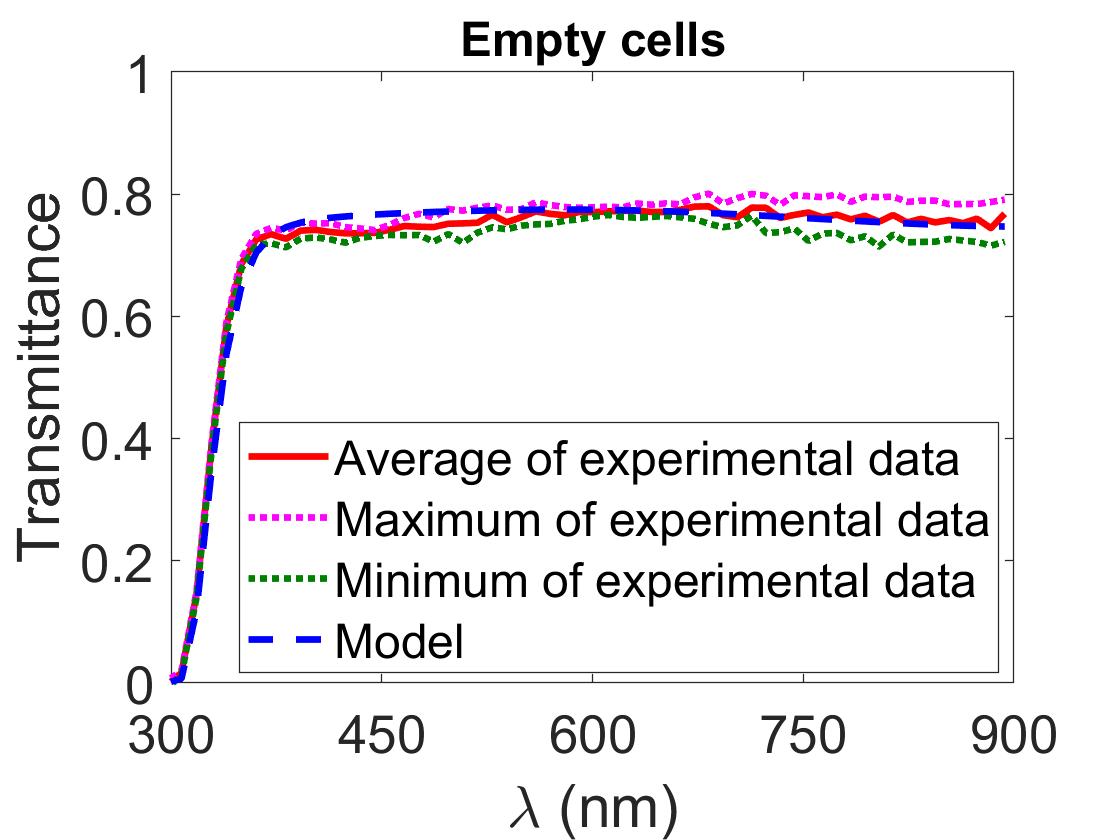}
\caption{Average of measured transmittance spectra of six empty
cells together calculated values using dielectric functions of substrate
layers with fitted parameters in Table \protect\ref{tab:glassITOPI}. The
solid red curve is the measured transmittance averaged over six cells, bounded
by the upper and lower extreme values (dotted curves) among the six measurements at
each wavelength. }
\label{fig:EC}
\end{figure}

The goodness of fit for the dielectric functions of the three substrate
layers was further tested by comparing the experimental measurements of
empty cells with the theoretical prediction of the transmittance of the
9-layer structure with $%
m=9,n_{1}=n_{5}=n_{9}=1,n_{2}=n_{8}=n_{glass}^{cG},n_{3}=n_{7}=n_{ITO}^{S},n_{4}=n_{6}=n_{PI}^{S}
$, using fitted model parameters in Table \ref{tab:glassITOPI}. The
experimental data is plotted as the average over the measurements of six
different empty cells, with upper and lower bounds from the six
measurements at each wavelength. Our theoretical prediction of the empty
cell shows reasonable agreement with these measurements, with a larger
discrepancy near $300-500$ nm as shown in Fig.~\ref{fig:EC}. This is mainly
due to the imperfection of the PI fitting at near-UV. We note that since the
fitting for the PI didn't show excellent agreement between theory and
measurement in the near-UV, the error in the PI fitting will carry over to
the fitting of the LC and subsequent dye-LC mixtures. Even though the
relative error in the near-UV made in the previous steps carried forward, it
is not significant enough to invalidate the subsequent steps. If a highly
accurate characterization in the near-UV is needed, more accurate
measurements will also be required.

\begin{table}[htb]
	\centering
	\begin{tabular}{|l|c|c|c|c|c|}
		\hline
		& $\varepsilon_\infty$ &  & $A_k$ & $\omega_{0k}$ & $\sigma_k$ (THz) \\ \hline
		Glass(Gaussian) & $2.3104$ & $k=1$ & $0.2107$ & $2\pi c/195.02nm$ & $34851$ \\ \hline
		&  & $k=2$ & $0.00001$ & $2\pi c/987.31nm$ & $3127$ \\ \hline
		ITO(Sellmeier) & $1.1221$ & $k=1$ & $1.4299$ & $2\pi c/275.01nm$ & $-$ \\ \hline
		&  & $k=2$ & $0.0641$ & $2\pi c/1000.01nm$ & $-$ \\ \hline
		PI (Sellmeier) & $2.4779$ & $k=1$ & $0.7831$ & $2\pi c/199.32nm$ & $-$ \\ \hline
		&  & $k=2$ & $0.5697 $ & $2\pi c/1315.22nm$ & $-$ \\ \hline
	\end{tabular}%
	\caption{Fitted parameters in the dielectric functions of glass, ITO and PI.
		Here $c$ is the speed of light in air.}
	\label{tab:glassITOPI}
\end{table}

\subsection{Dielectric functions of dye-doped liquid crystal mixtures}

\label{sec_LC_dye}

Liquid crystals are typically birefringent, as are oriented dichroic dyes.
When the transmittance of a mixture is measured with the polarization of
incident light either parallel or perpendicular to the LC director, only one
principal refractive index contributes to the transmittance. The alignment
of LCs and dye mixtures was confirmed by applying a voltage across the cell.
Since the LC has a negative dielectric anisotropy at low frequency, no
change in orientation is expected, and none was observed.

We assume that dielectric functions of the LC and of the dye can be treated as
additive. This is valid when the mixture is homogeneous and the
concentration of dye is low. The experimental data suggest that the
locations of the absorption peaks and their broadenings are nearly the same
for both polarizations and only the magnitudes differ. The two principal
dielectric functions of dye-doped LC mixtures are then given as 
\begin{eqnarray}
\varepsilon _{\parallel }^{LC+dye} &=&\varepsilon _{\infty \parallel
}^{LC}+\sum_{k=1}^{2}\chi _{k}^{S}(\omega ;A_{k\parallel }^{LC},\omega
_{0k}^{LC})+c_{dye}\sum_{k=1}^{N_{dye}}\chi
_{k}^{cG}(\omega ;A_{k\parallel }^{dye},\omega _{0k}^{dye},\sigma
_{k}^{dye}),  \label{eqn_dye_a} \\
\varepsilon _{\perp }^{LC+dye} &=&\varepsilon _{\infty \perp
}^{LC}+\sum_{k=1}^{2}\chi _{k}^{S}(\omega ;A_{k\perp }^{LC},\omega
_{0k}^{LC})+c_{dye}\sum_{k=1}^{N_{dye}}\chi
_{k}^{cG}(\omega ;A_{k\perp }^{dye},\omega _{0k}^{dye},\sigma _{k}^{dye}),
\label{eqn_dye_b}
\end{eqnarray}%
where $c_{dye}$ denotes the concentration of the dye, and $N_{dye}$ the
number of oscillators associated with the dye. The first sum accounts for
the contribution from the LC host and the second from the dye
guest. In Eqs.~\eqref{AvgRef}-\eqref{AvgTrans}, we set $%
m=9,n_{1}=n_{9}=1,n_{2}=n_{8}=n_{glass}^{cG},n_{3}=n_{7}=n_{ITO}^{S},n_{4}=n_{6}=n_{PI}^{S}
$, and try to find $n_{5}$ which minimizes the sum of squared errors of the
model and data. In what follows, we first obtain the pure LC contribution of
the first two terms before the dye is introduced.

\subsubsection{liquid crystal host}

\label{Sect_LCs}

The ordinary refractive index $n_{o}$ of the LC typically ranges from $%
1.50-1.57$ while extraordinary refractive index $n_{e}$ can range from $%
1.5-1.9$ depending on molecular structure, temperature and wavelength \cite%
{Li2005_2}. Many models, Cauchy or Sellmeier, have been proposed to
model the wavelength and temperature dependence of the refractive indices of
LCs \cite{Li2004, Abdul1991,Wu1991}. 

\begin{figure}[htb]
\centering
(a)\includegraphics[width=.4\linewidth]{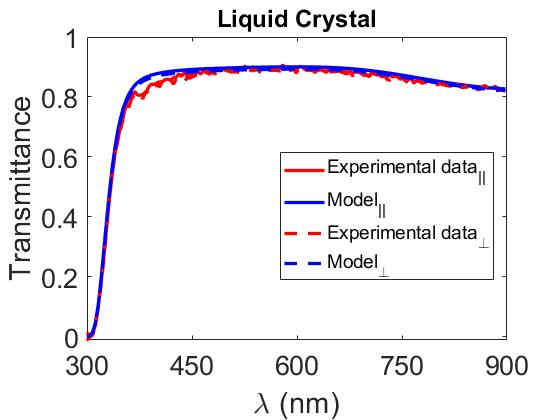} \quad(b) %
\includegraphics[width=.4\linewidth]{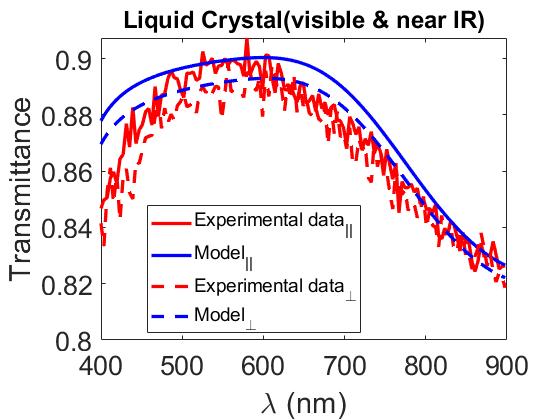}
\caption{(a) Transmittance spectra of an LC sample without dye in a cell. (b)
Expanded region of (a) in the visible region. Solid red curves show experimental
results, and dashed blue curves indicate predictions from model with two Sellmeier oscillators for ${\boldsymbol{%
\protect\varepsilon}}_{LC}^{S} $ with fitted parameters in Table \protect
\ref{tab:PureLCA_dye}.}
\label{fig:PureLCA}
\end{figure}

\begin{figure}[htb]
\centering
(a)\includegraphics[width=.4\linewidth]{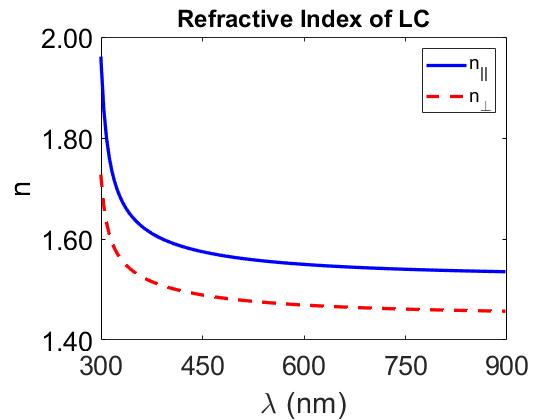} 
\caption{Refractive indices of nematic LC for polarization (a) parallel
and perpendicular to the LC director.}
\label{fig:PureLCARI}
\end{figure}

In this work, the characterization of the dielectric function of the LC host
is based on the transmittance spectra of the cell containing pure LC host
from $300-900$ nm. The LC appears to be transparent at visible wavelengths, but with appreciable reduction of the transmittance in the near-UV. In order to understand the origin of this reduction, we carried out a further set of experiments, where three pure LC cells with the same glass substrates but with different thicknesses, $5\mu m, 20 \mu m, 47 \mu m$, were prepared and transmittance spectra were measured. The results show that the transmittances of the LC cells are apparently independent of gap thickness, which implies that the LC host is nonabsorptive. This prompts us to use the Sellmeier equation for the dielectric function of LC host. A two-term Sellmeier equation with one peak located at $125.01$ nm and the other at $289.62$nm was obtained from the fit. Fig.~\ref{fig:PureLCA}%
	(a) shows reasonable agreement between the theoretical predictions and
	experimental measurements for both polarizations; it is expanded in the
	visible range in Fig.~\ref{fig:PureLCA}(b) to allow better visual
	assessment.  The average relative error is about $1\%$ in the range of $400-900$ nm. The larger discrepancy of fitting and data occurs at short wavelengths in the near-UV, where larger experimental measurement errors likely occur.  Figure~\ref{fig:PureLCARI} shows the two principal refractive
	indices of the host LC calculated with Sellmeier with
	parameters given in Table~\ref{tab:PureLCA_dye}. Our LC host is a multicomponent mixture, with manufacturer provided refractive indices $n_e=1.55$ and $n_0=1.47$ without any wavelength dependence information. Our results show reasonable agreement with the manufacture's data.

\subsubsection{dichroic dye and liquid crystal mixtures}

Lastly, we report the characterization of the dielectric functions of two
dye-LC guest-host mixtures. The concentration of dye in the dye-LC mixture
is approximately $1\%$. The dielectric functions of the dye-LC mixtures are
modeled as Eqs.~\eqref{eqn_dye_a} -\eqref{eqn_dye_b}. The first two terms in
both equations have already been acquired in Sect.~\ref{Sect_LCs}; here we
only need to obtain the last term which only contains the dye contribution.

\begin{figure}[h]
\centering
(a)\includegraphics[width=.4\linewidth]{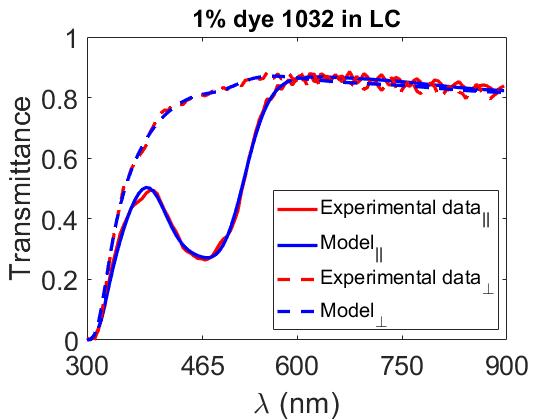} \quad(b)%
\includegraphics[width=.4\linewidth]{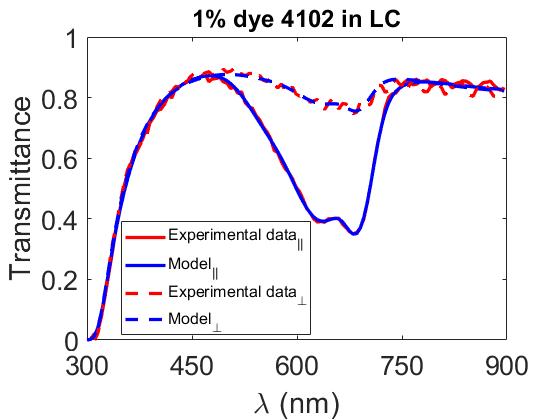}
\caption{Transmittance spectra of mixture (a) $1\%$ dye 1032 in LC and (b) $1\%$ dye
4102 in LC. Solid red curves are experimental data, and dashed blue curves are
calculated transmittance of structures using the combined Sellmeier and causal Gaussian
oscillator model for ${\boldsymbol{\protect\varepsilon}}_{LC+dye}$ with
fitted parameters in Table \protect\ref{tab:PureLCA_dye}. }
\label{fig:1032_4102}
\end{figure}

\begin{figure}[h]
\centering
(a)\includegraphics[width=.4\linewidth]{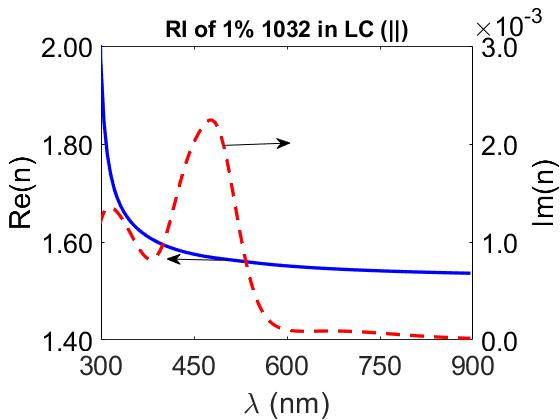} \quad(b)%
\includegraphics[width=.4\linewidth]{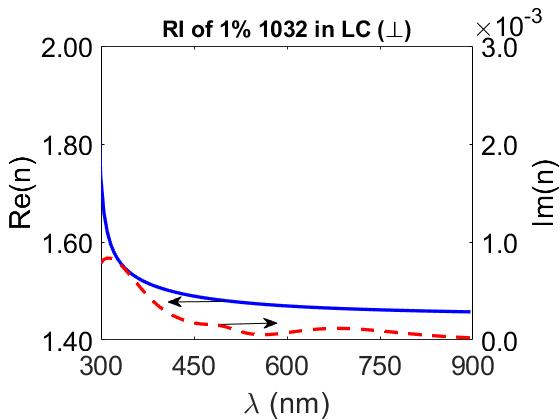} (c)%
\includegraphics[width=.4\linewidth]{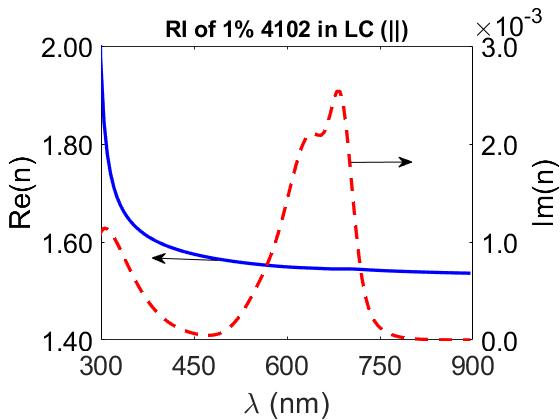} \quad (d)%
\includegraphics[width=.4\linewidth]{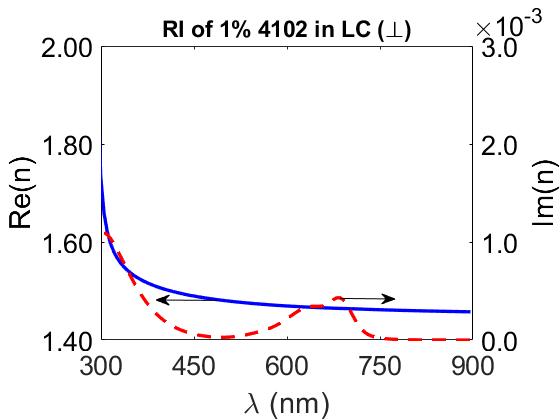}
\caption{Top: Refractive indices of $1\%$ dye 1032 in LC. Bottom: Refractive indices of $1\%$
dye 4102 in LC. Left: principle refractive indices for polarization along the LC director;
Right: principle refractive indices for polarization perpendicular to LC
director.}
\label{fig:RI1032_4102}
\end{figure}

The first dye, referred to as dye 1032, is orange in color. We have used
four causal Gaussian oscillators for the contribution of this dye. One is
peaked at $304.07$ nm for the slight absorption near $300$ nm, and two are at $455.08$ and $495.89$nm for the strong absorption near $465$nm, and one at $683.51$ nm for slight absorption at longer wavelength. We note that a single
oscillator near the $465$nm is not sufficient to accurately fit both the
breadth and shape of the transmittance spectrum. The second dye, referred to
as dye 4102, is blue in color. The transmittance spectrum clearly shows two
peaks between $600-750$ nm, as shown in Fig.~\ref{fig:1032_4102}(b). We have
tried to use three causal Gaussian oscillators with one in the near-UV and
two in the visible, but these can't reproduce the shape of transmittance
spectrum near $600$ nm. We have then used four causal Gaussian oscillators,
one is peaked at $291.23$ nm for the absorption in the near-UV, and three in the
visible at $594.28$, $645.31$ and $688.63$ nm, respectively. The fit results for LC
mixture with dyes 1032 and 4102 are shown in Fig.~\ref{fig:1032_4102}(a) and
(b), respectively, showing excellent agreement of the experimental
measurements and theoretical predictions. The average relative errors are $2\%$ in the range of 400-900 nm. Figure \ref{fig:RI1032_4102} shows
the corresponding refractive indices of dye-LC mixtures calculated from the combined Sellmeier and
causal Gaussian oscillator models with parameters given in Table \ref%
{tab:PureLCA_dye}. Both dyes show only weak absorption for perpendicular
polarization in the visible;  the values of $A_{\perp k}$ are much
smaller than those of $A_{\parallel k}$. This is an indication of a good
alignment of dye molecules with the nematic director.

\begin{table}[htb]
	\centering
	\begin{tabular}{|c|c|c|c|c|}
		\hline
		& $\varepsilon_{\infty\parallel}=1.1649  $ & $\varepsilon_{\infty \perp}=1.0149$
		&  &  \\ \hline
		liquid crystal & $A_{\parallel k}$ & $A_{\perp k}$ & $\omega_{0k}$ & $\sigma_k$ (THz) \\ \hline
		$k=1$ & 1.0671  & 1.0341 & $2\pi c/125.01nm$ & $-$ \\ \hline
		$k=2$ & 0.0932 & 0.0482 & $2\pi c/289.62nm$ & $-$ \\ 
		\hhline{|=|=|=|=|=|} \text{dye 1032} &  &  &  &  \\ \hline
		$k=1$ & 0.5211 & 0.2701 & $2\pi c/304.07nm$ & $12896$ \\ \hline
		$k=2$ & 0.5098 & 0.0001 & $2\pi c/455.08nm$ & $16593$ \\ \hline
		$k=3$ & 0.3207 & 0.0210 & $2\pi c/495.89nm$ & $25654$ \\ \hline
		$k=4$ & 0.0191 & 0.0298 & $2\pi c/683.51nm$ & $7236$ \\ 
		\hhline{|=|=|=|=|=|} \text{dye 4102} &  &  &  &  \\ \hline
		$k=1$ & 0.4001 & 0.3799 & $2\pi c/291.23nm$ & $15375$ \\ \hline
		$k=2$ & 0.2234 & 0.0288 & $2\pi c/594.28nm$ & $12708$ \\ \hline
		$k=3$ & 0.5089 & 0.0794 & $2\pi c/645.31nm$ & $18050$ \\ \hline
		$k=4$ & 0.4119 & 0.0686 & $2\pi c/688.63nm$ & $40319$ \\  \hline
	\end{tabular}%
	\caption{Fitted parameters in the combined Sellmeier and causal Gaussian oscillator model with
		contributions from liquid crystal, dye 1032 and dye 4102. Here $c$ is the speed
		of light in air. }
	\label{tab:PureLCA_dye}
\end{table}

\begin{figure}[htb]
\centering
(a)\includegraphics[width=.4\linewidth]{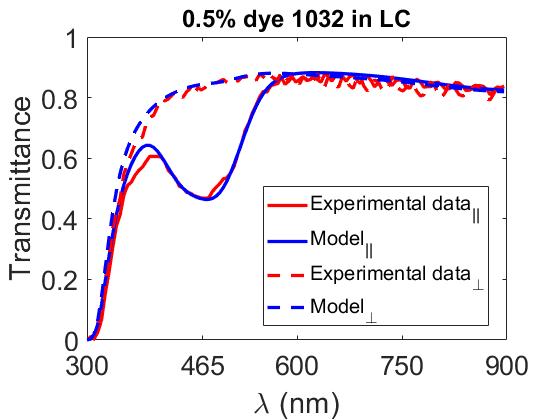} \quad (b) %
\includegraphics[width=.4\linewidth]{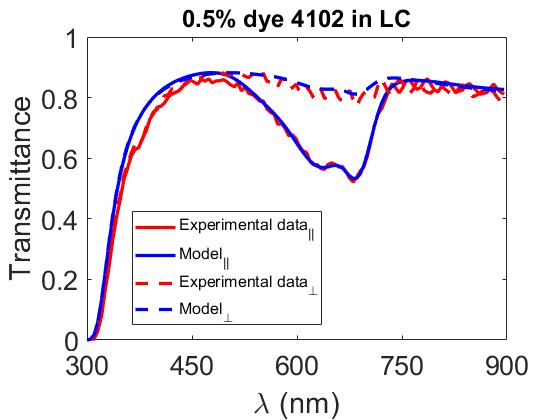}
\caption{Transmittance spectra of $0.5\%$ dye (a) 1032 and (b) 4102 in LC.
Solid red curves are experimental data, and dashed blue curves are calculated values using the combined Sellmeier and causal Gaussian oscillator model
for ${\boldsymbol{\protect\varepsilon}_{dye+LC}}$ with fitted
parameters in Table \protect\ref{tab:PureLCA_dye} and scaled amplitudes.}
\label{fig:half}
\end{figure}

To test the usefulness of our model, we examine the goodness of the
theoretical prediction of $0.5\%$ dye concentration in LC. We set $%
c_{dye}=0.5\%$ in Eqs.~\eqref{eqn_dye_a}$-$\eqref{eqn_dye_b} and keep all the
other parameters the same to obtain the dielectric functions of $0.5\%$
dye-LC mixtures via Eqs.~\eqref{eqn_dye_a}$-$\eqref{eqn_dye_b}. We then use
Eq.~\eqref{AvgTrans} with $m=9$ to calculate the transmittance spectra of
the cell. The results are given in Fig.~\ref{fig:half}, showing excellent
agreement with experimental measurements $0.5\%$ dye in LC,  \textcolor{black}{with average relative errors  $3\%$ in the range of 400-900 nm.} We are confident
to conclude that the dye and liquid crystal mixtures we have studied behave
linearly, which makes the modeling straightforward and we can effectively
predict the transmittance at any concentration of dye-LC mixture, in the low
concentration regime.

\textcolor{black}{In Ref.~\cite{Goda2013}, the assumed dielectric function has uncorrelated real and imaginary parts. This allows the use of the Beer-Lambert law to obtain the extinction coefficients in a simple way, disregarding their dependence on the substrate layers.   Our method, however, deals with the more general case when the real and imaginary parts cannot be assumed to be independent and multiple reflections are properly considered in all the layers. }

\section{Conclusions}

\label{Sec_Con} In this work, we have characterized the wavelength
dependence of the refractive indices of dichroic dye-liquid crystal
guest-host systems from the near-UV/vis ($300$ to $900$ nm) transmittance
spectra. The dye-LC mixtures are contained in planar cells, where the
substrates are glass coated with ITO and PI layers. The PI layer is buffed
in the plane to impose in-plane orientation of the LC director, which in
turn aligns the dichroic dye in the same direction. When linearly polarized
light parallel or perpendicular to the LC director is normally incident on
the sample, only one principal refractive index of the dye-LC system
contributes to the transmittance of the sample. These experimental
arrangments allow characterization of the two principal components of a
uniaxial material independently.

We have used the Sellmeier equation for the dielectric functions of the
transparent layers away from their absorption peaks, and the causal Gaussian
oscillator model for layers showing appreciable absorption in the
near-UV/vis. Both models satisfy the Kramers-Kronig causality relations. The
parameters in the dielectric function of each layer were obtained by
minimizing the sum of squared errors between theoretical predictions and
measurements of the transmittance.

Experimental measurements on substrate layers show no polarization
dependence, hence all substrate layers are modeled as optically isotropic.
The dielectric function of glass is modeled by the causal Gaussian
oscillators with one peak in the near-UV for the strong absorption around $%
300$ nm and one in the near IR for the slight decrease in the transmittance
at long wavelengths. The dielectric functions of the ITO and PI layers are
modeled by Sellmeier model with two oscillators: one near-UV and one near-IR.

To model the two principal dielectric functions of the dye-LC mixtures, we
have used the combined Sellmeier and causal Gaussian oscillator models. We assume that each
oscillator has the same resonance location and width, but different
amplitudes, in the two principal directions, as indicated by the
experimental data. The dielectric function of the liquid crystal host was
characterized first, before introduction of the dye. The dielectric function of the pure LC host is modelled by the Sellmeier equations with two near-UV oscillators. The dye contribution to the dielectric function of dye-LC
mixtures is modeled by the addition of causal Gaussian oscillators.  For the two dye-LC mixtures reported, we have used four
oscillators for each dye.
Each dye contains one oscillator in the near-UV, and the others are placed at
visible wavelengths to account for the absorption.

The causal Gaussian oscillator model can accurately characterize the
dielectric function of dichroic dyes in liquid crystals in the near-UV/vis
range in the low concentration regime. We have also used the same approach
to model commercial dyes in isotropic solvents and obtained similar
excellent quantitative agreement between theory and experiment. 

Knowing the dielectric functions of dichroic dyes in LCs allows further
analysis of different variation of the guest-host LC devices. Based on our
results we expect to be able to predict the transmittance and reflectance of
samples with inhomogeneous director fields, with different dye concentration
and arbitrary angles of incidence in similar multilayer structures.
The transmittance spectrum of the sample can be mapped onto the colors
perceived by the human eye, leading, ultimately to the optimal design of LC
devices. 

\section*{Acknowledgments}

The dye-mixtures are provided by AlphaMicron Inc. This work was supported by Air Force contract FA8649-20-C-0011 as part of the STTR AF18B-T003 Electronically Dimmable Eye Protection Devices (EDEPD) program and by the Office of Naval
Research through the MURI on Photomechanical Material Systems (ONR
N00014-18-1-2624).

\section*{Conflicts of interest}

There are no conflicts of interest to declare.

\end{document}